\documentclass[runningheads]{llncs}
\usepackage{graphicx}
\begin{document}
%
\title{Automated head and neck tumor segmentation from 3D PET/CT \protect\\ HECKTOR 2022 challenge report}
%
%
\author{Andriy Myronenko \and 
Md Mahfuzur Rahman Siddiquee \and
Dong Yang \and
Yufan He \and
Daguang Xu 
}
\authorrunning{A. Myronenko et al.}
%
\institute{NVIDIA, Santa Clara, CA \\
\email{\{amyronenko,mdmahfuzurr,dongy,yufanh,daguangx\}@nvidia.com}}
\maketitle              
\begin{abstract}
Head and neck tumor segmentation challenge (HECKTOR) 2022 offers a platform for researchers to compare their solutions to segmentation of tumors and lymph nodes from 3D CT and PET images. In this work, we describe our solution  to HECKTOR 2022 segmentation task. 
We re-sample all images to a common resolution,  crop around head and neck region, and train SegResNet semantic segmentation network from MONAI. We use 5-fold cross validation to select best model checkpoints. The final submission is an ensemble of 15 models from 3 runs. Our solution (team name NVAUTO) achieves the 1st place on the HECKTOR22 challenge leaderboard with an aggregated dice score of 0.78802\footnote{https://hecktor.grand-challenge.org/evaluation/segmentation/leaderboard/}. It is implemented  with Auto3DSeg\footnote{https://monai.io/apps/auto3dseg}.


\end{abstract}

\section{Method}
\subsection{Introduction}

Head and Neck (H\&N) cancer is the fifth most prevalent cancer type globally by incidence rate~\cite{hecktor22_B}. Specialized medication and radiotherapy are a standard treatment types, but cancer recurrences occur in almost a half of the cases within the first years after treatments. 3D medical imaging, such as  Computer Tomography (CT) and Positron Emission Tomography (PET), provides insights into disease prognosis and treatment planning. 

Head and neck tumor segmentation challenge (HECKTOR) provides an opportunity for researchers to develop 3D algorithms for the segmentation of H\&N primary tumors (GTVp) in 3D PET/CT scans. HECKTOR 2022~\cite{hecktor22_A,hecktor22_B} is a third edition of the challenge which consists of 883 cases (524 labeled cases were provided for training), each with 3D CT, 3D PET rigidly registered to a common frame, but at different resolutions. The ground truth 3D labels provide dense 3D annotations of 2 structures:  gross tumor volumes of the primary tumors (GTVp) and lymph nodes (GTVn). Generally PET images highlight tumor activity at a lower resolution, whereas CT images provide higher resolution anatomical details. In case of the radiotherapy treatment, the tumor delineation must be done in the CT coordinate system, which which will be used to calculate the radiation dose to the tumor region.  The HECKTOR22 challenge also includes the second task of outcome prediction, but here we focus solely on the segmentation task.  The data used in this challenge comes from multiple institutions  (9 centers in total), including 4 centers in Canada, 2 centers in Switzerland, 2 centers in France, and 1 center in the United States for a total of 883 patients with annotated GTVp and GTVn~\cite{hecktor22_A,hecktor22_B}.

\begin{figure}[!ht]
    \centering
    \includegraphics[width=0.49\textwidth]{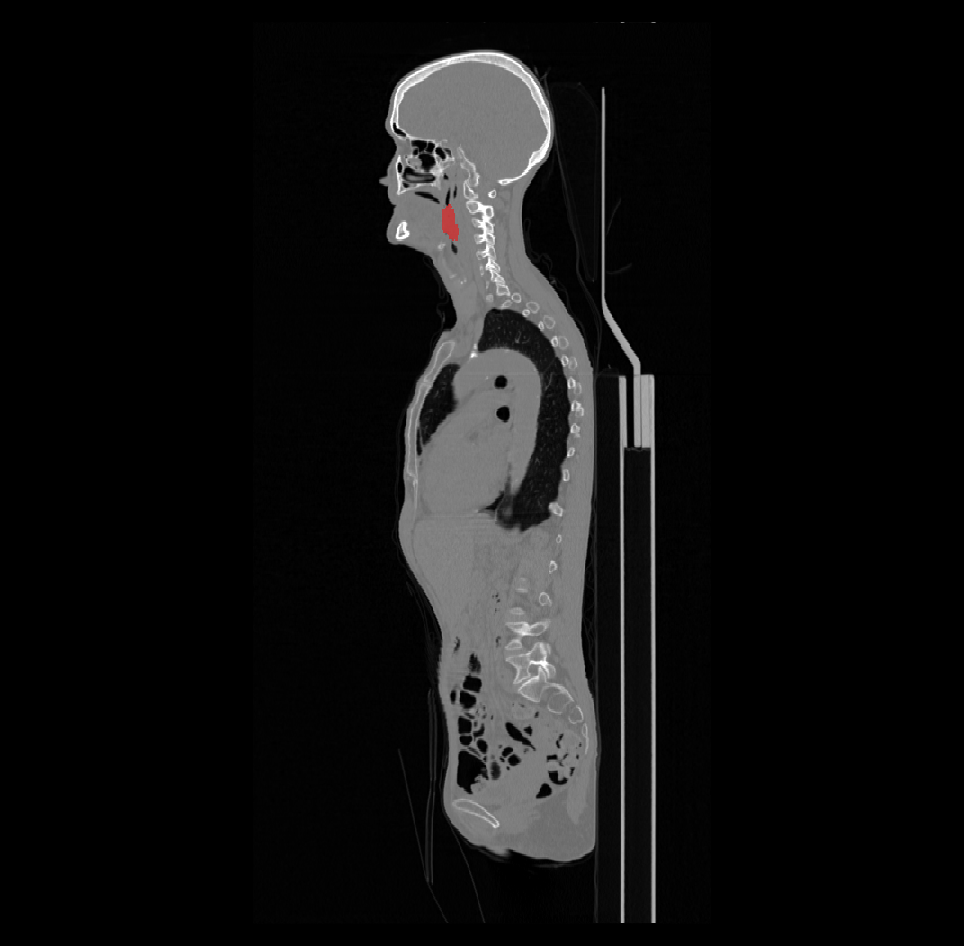}
    \includegraphics[width=0.49\textwidth]{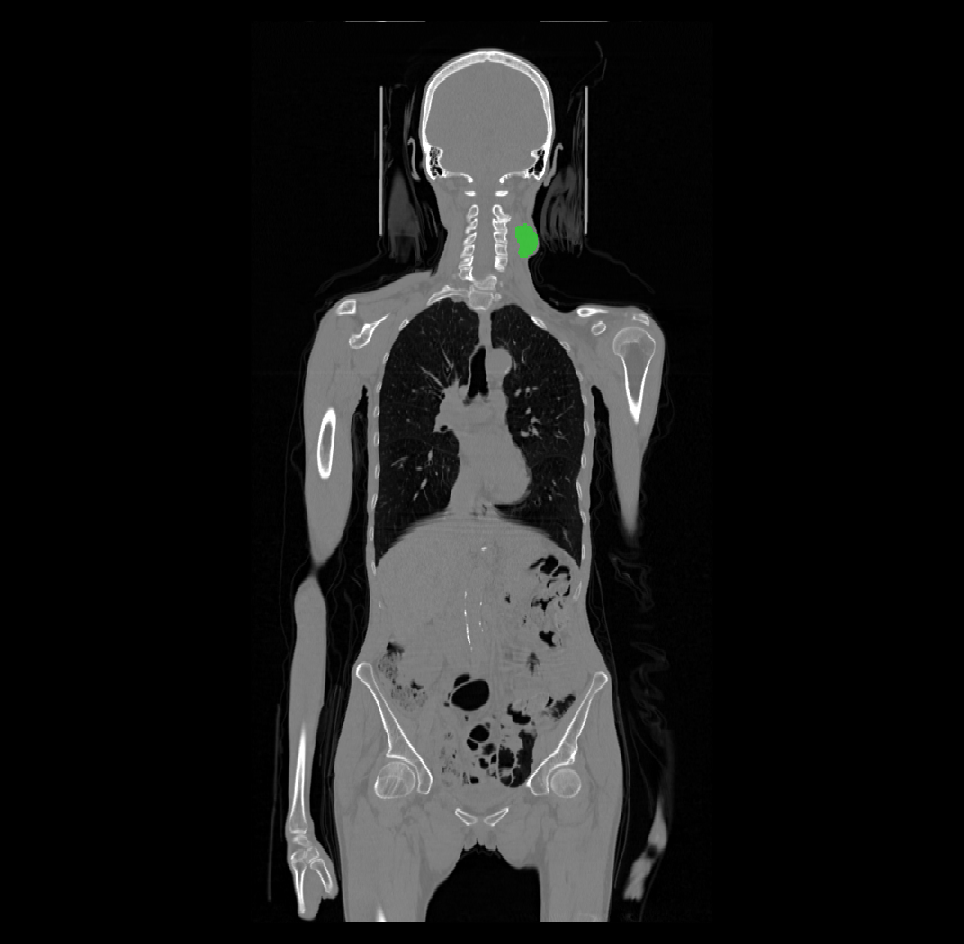}
    \includegraphics[width=0.49\textwidth]{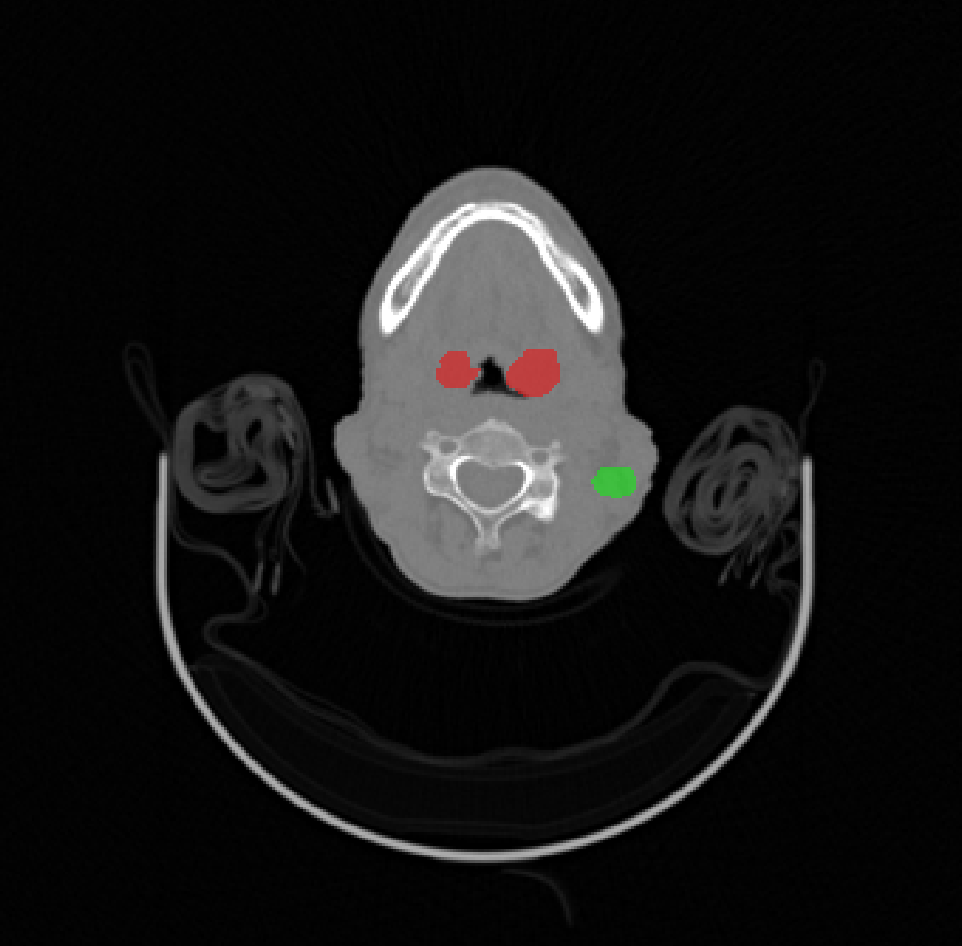}
    \includegraphics[width=0.49\textwidth]{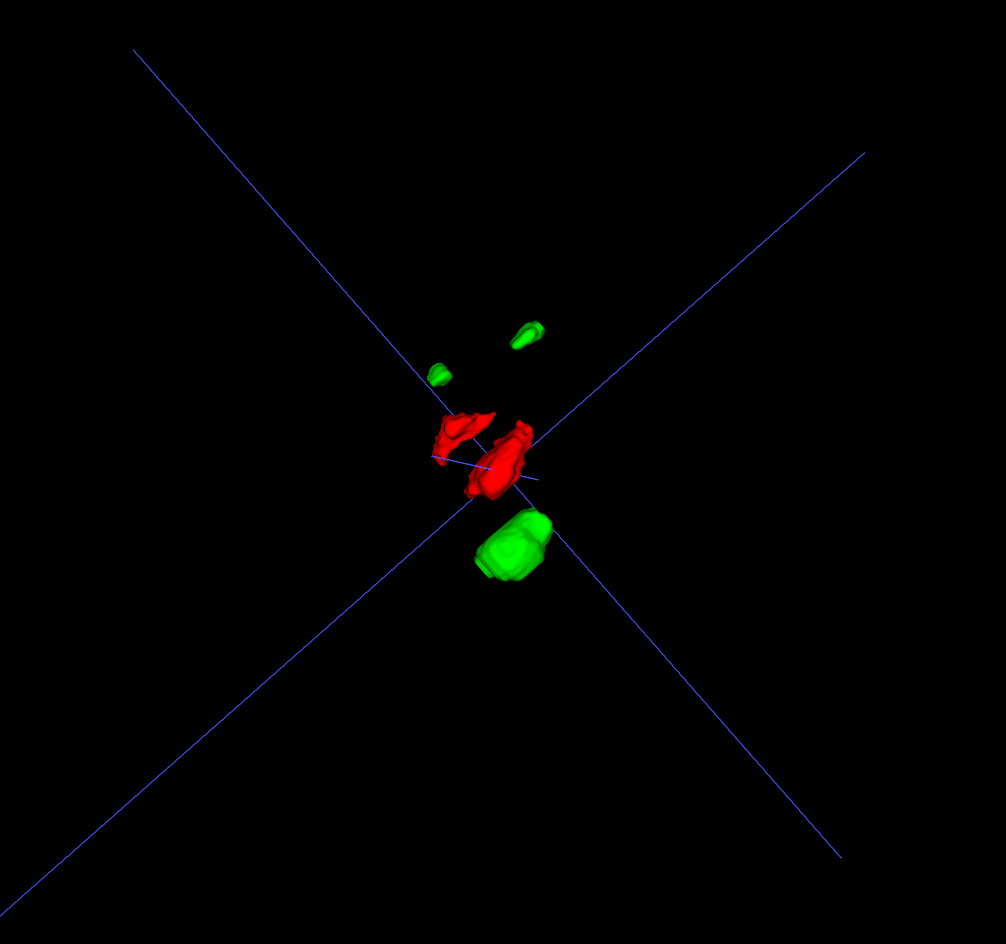}
    \caption{An example of CT image showing sagital, coronal and axial slices with tumors (in red) and lymph nodes (in green) mask overlays. A 3D visualization demonstrates that the tumor consists of 2 components around the neck (in red) and the lymph nodes region has 3 components annotated (in green). The CT size is 500x500x978mm}
    \label{fig:example1}
\end{figure}

The training dataset with the ground truth labels consists of  524 cases with average 3D CT size of 512x512x200 voxels at 0.98x0.98x3 mm average resolution, and with average 3D PET size of 200x200x200 voxels  at 4x4x4 mm. The CT and PET image pairs where rigidly aligned to a common origin, but remain at different sizes and resolutions.  Many cases provided were almost a full body CT/PET pairs. This provides both computational and algorithmic challenge, since the imaging region is as large as  500x500x1000 mm of the body anatomy, whereas the tumor region covers less then 5\% of the input images. 

\begin{figure}[!t]
    \centering
    \includegraphics[width=0.49\textwidth]{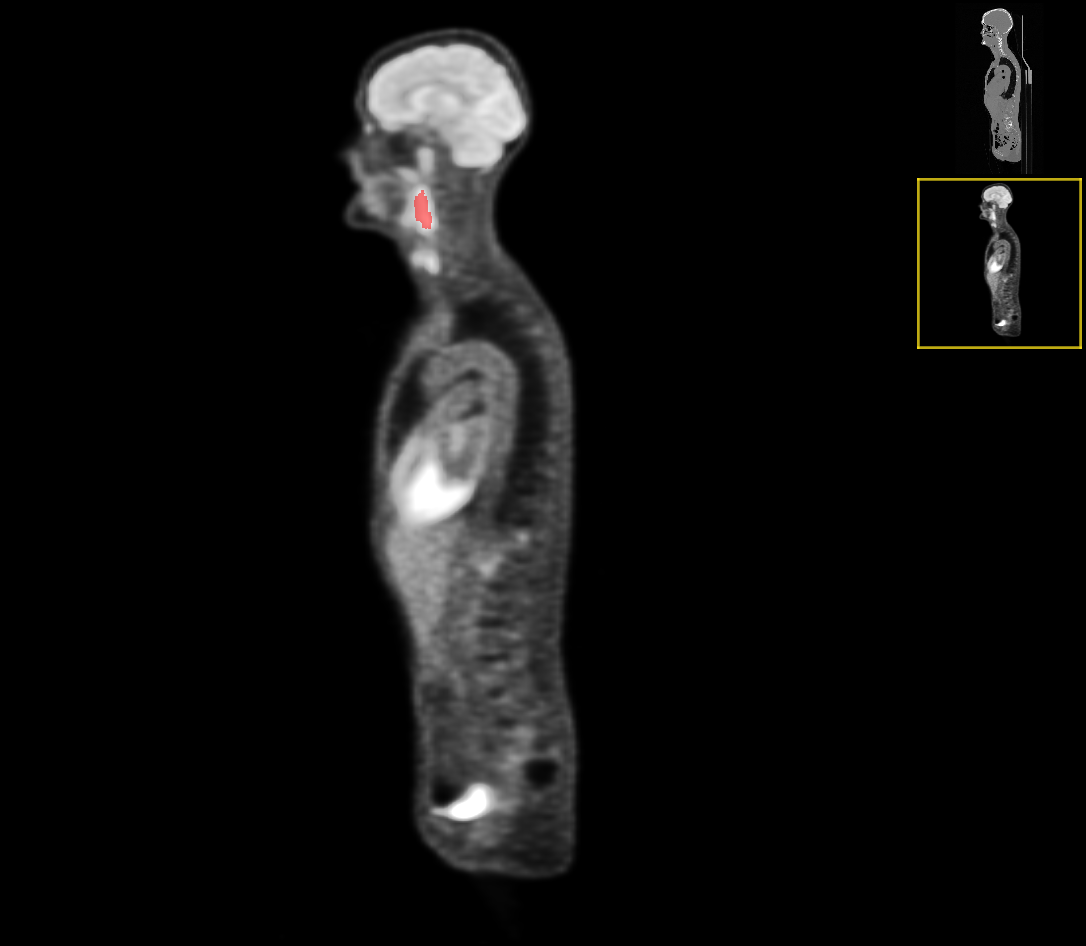}
    \includegraphics[width=0.49\textwidth]{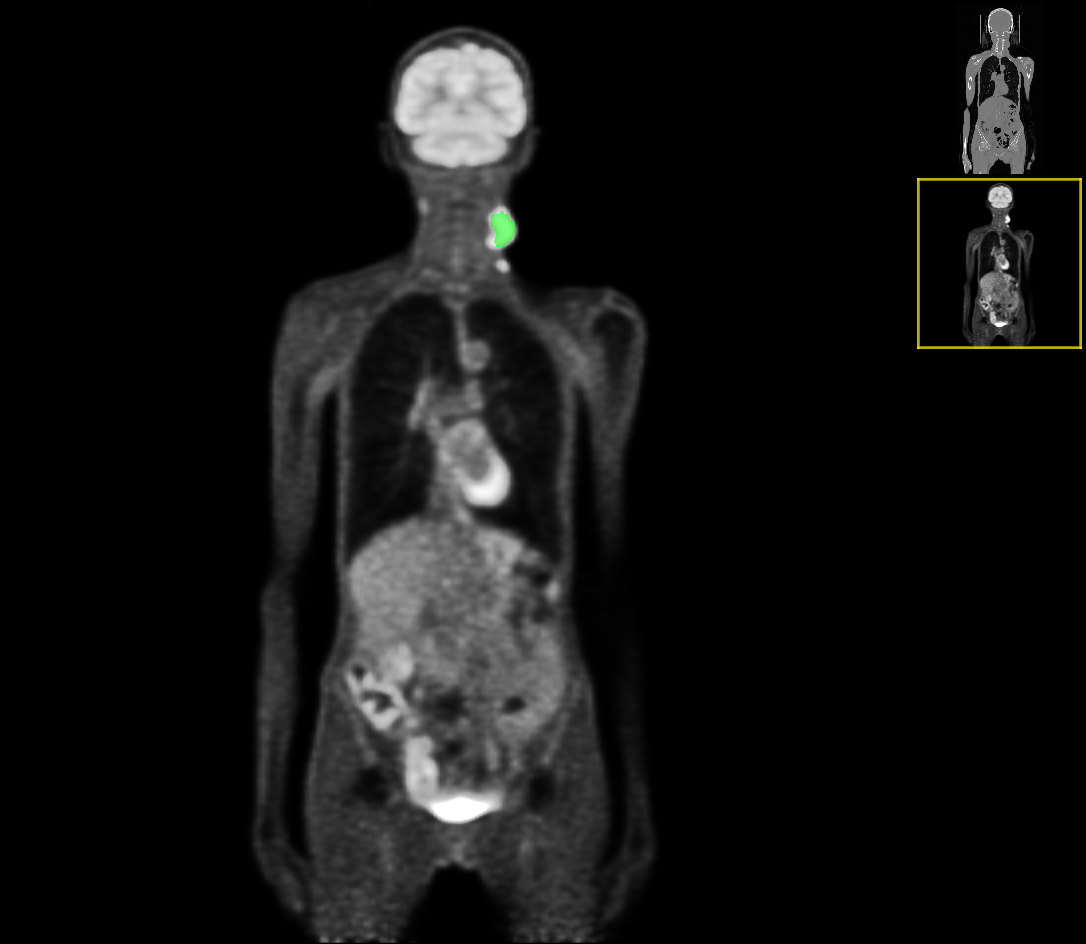}
    \caption{An example of PET  image showing sagital and coronal slices with tumors (in red) and lymph nodes (in green) mask overlays.}
    \label{fig:example2}
\end{figure}

The ground truth labels usually include a single mass of the primary tumor (but in some cases it was absent completely or had two components), and several connected components of the annotated lymph nodes.  An example case of CT and the corresponding PET image with ground-truth overlays is shown in Figures~\ref{fig:example1} and ~\ref{fig:example2}.

\subsection{Method}

\begin{figure}[t]
    \centering
    \includegraphics[width=0.8\textwidth]{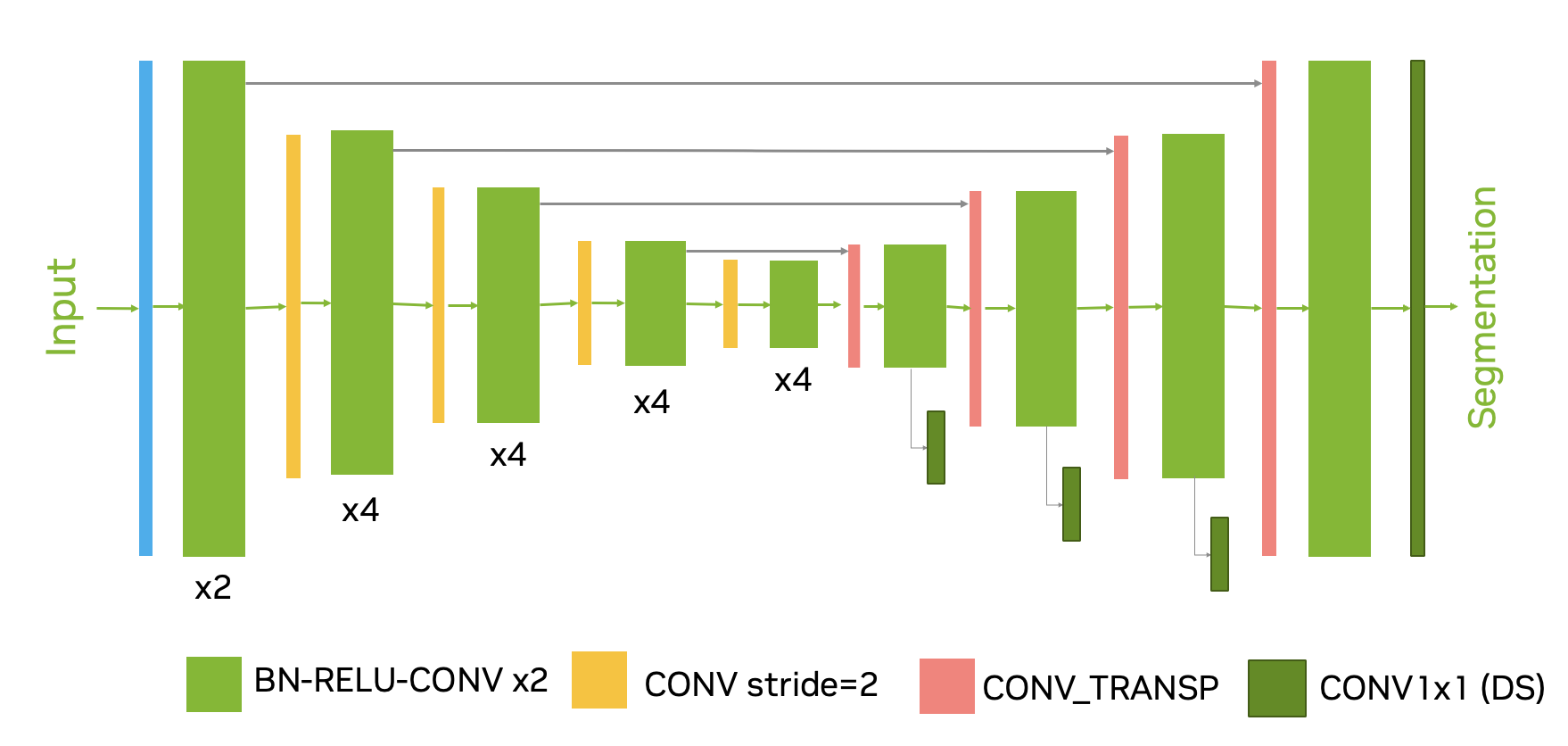}

    \caption{SegResNet network configuration. The network uses repeated ResNet blocks with batch normalization and deep supervision}
    \label{fig:net}
\end{figure}

We implemented our approach with MONAI\footnote{https://github.com/Project-MONAI/MONAI}~\cite{monai}, we used Auto3DSeg\footnote{https://monai.io/apps/auto3dseg} system to automate most parameter choices. For the main network architecture we used SegResNet\footnote{https://docs.monai.io/en/stable/networks.html\#segresnet}, which is an encode-decoder based semantic segmentation network based on~\cite{myronenko20183d}, with deep supervision (see Figure~\ref{fig:net}).


Overall, our approach consists of the following steps: data analysis to determine appropriate image normalization parameters and tumor regions, image re-sampling and training of several runs using 5-fold cross-validation, and finally model ensembling. 

\subsubsection{Data preparation} We resample both CT and PET input images to the same size and  1x1x1mm isotropic resolution, and crop an approximate region around head and neck. The steps to crop an approximate H\&N region are basic and based on a relative anatomy position within the PET/CT images:
\begin{itemize}
	\item Detect the top of the head (top of the bounding box), based on a simple PET thresholding
	\item Detect the H\&N center-line (xy coordinate) based on the average foreground of top slices  
	\item Crop the bounding box of 200x200x310mm centered on the center-line. 
\end{itemize}
This simple approach had 100\% success rate on the training set to cover  the H\&N region fully.  We contemplated a more sophisticated approach based on deep-learning, but it was not necessary in this case. 

Cropping the approximate region is the first step both during training and during inference.  During training it significantly reduces the input image size (e.g. from 500x500x900 to 200x200x310 voxels), which speeds up training and avoids unnecessary network strain to differentiate other anatomies (e.g. in abdominal region). 

\subsubsection{Data normalization}

We re-scale input CT image intensity from a predefined range to 0..1 interval, as determined by data analysis to include the intensity pattern variations within the foreground regions, followed by a sigmoid. For PET image, we normalize it to zero mean and standard deviation one, followed by a sigmoid. Sigmoid function here is used as an alternative to hard intensity clamping. After normalization, input images are concatenated to form a 2 channel input image.

\subsubsection{Model} For the model, we used the encoder-decoder semantic segmentation model SegResNet from MONAI based on~\cite{myronenko20183d} with deep supervision. The encoder part uses ResNet~\cite{he2016identity} blocks, and includes 6 stages of 1, 2, 2, 4, 4, 4  blocks respectively. We follow a common CNN approach to downsize image dimensions by 2 progressively and simultaneously increase feature size by 2.  All convolutions are 3x3x3 with an initial number of filters equal to 32. The encoder is trained with $192\times192\times192$ input region.  The decoder structure is similar to the encoder one, but with a single block per each spatial level. Each decoder level begins with upsizing with transposed convolution: reducing the number of features by a factor of 2  and doubling the spatial dimension, followed by the addition of encoder output of the equivalent spatial level. The end of the decoder has the same spatial size as the original image, and the number of features equal to the initial input feature size, followed by a 1x1x1 convolution into 3 channels and a softmax (a background and two foreground classes).

\section{Training Method}

\subsection{Dataset}

We use the HECKTOR22 dataset~\cite{hecktor22_A,hecktor22_B} . We randomly split the entire dataset into 5 folds and trained a model for each fold. We did not use any additional data or pre-trained models, and we did not use any of the meta-data information (such patients gender or age) provided by the organizers. 

\subsection{Cropping}

We crop a random patch of 192x192x192 voxels from the H\&N extracted area centered on the foreground classes with probabilities of 0.45  for tumor and 0.45 for lymph nodes (and 0.1 for background). 

\subsection{Augmentations}

We use random Affine and Flip augmentations followed by intensity augmentations for CT channel only. The CT augmentations include random intensity scale, shift, noise and blurring.

\subsection{Loss} We use the combined Dice + CrossEntropy loss. The same loss is summed over all deep-supervision sublevels:

\begin{equation}
Loss= \sum_{i=0}^{4} \frac{1}{2^{i}} Loss(pred,target^{\downarrow}) 
\end{equation}
where the weight $\frac{1}{2^{i}}$ is smaller for each sublevel (smaller image size) $i$. The target labels are downsized (if necessary) to match the corresponding output size using nearest neighbor interpolation
\subsection{Optimization}

We use the AdamW optimizer with an initial learning rate of $2e^{-4}$ and decrease it to zero at the end of the final epoch using the Cosine annealing scheduler. All the models were trained for 300 epochs with deep supervision. We use batch size of 1 per GPU, and train on 8 GPUs 16Gb NVIDIA V100 DGX machine (which is equivalent to batch size of 8). We use weight decay regularization of $1e^{-5}$.

\section{Results}

Based on our data splits,  a single run 5-folds cross-validation results are shown in Table~\ref{tab:result}.  On average, we achieve $0.7989$ cross-validation performance in terms of aggregated Dice metric. 

\begin{table}[h!]
    \centering
    \begin{tabular}{| c | c | c | c | c | c |}
        \hline
        {\textbf{Fold 1}} & {\textbf{Fold 2}} & {\textbf{Fold 3}} & {\textbf{Fold 4}} & {\textbf{Fold 5}} & {\textbf{Average}} \\
        \hline
        0.7933 &	0.7862 &	0.7816 &	0.8275 &	0.8059 & 0.7989 \\

        \hline
    \end{tabular}
    \caption{Aggregated dice metric using 5-fold cross-validation.}
    \label{tab:result}
\end{table}

For the final submission we use 15 models total, 3 fully trained runs. The challenge allowed only 3 submissions total, and required to submit dense prediction masks for 359 cases (saved in CT size/resolution). Our results are in Table~\ref{tab:result2}. All 3 of our submission are the top 3 submissions on HECTOR22 challenge leaderboard\footnote{https://hecktor.grand-challenge.org/evaluation/segmentation/leaderboard/}.
 
\begin{table}[h!]
    \centering
    \begin{tabular}{| l | l | c | c | c |}
        \hline
        \textbf{submission} & \textbf{note} & \textbf{tumor} & \textbf{lymph nodes} & \textbf{Total} \\
        \hline
        One & ensemble mean & 0.78797 & 0.77468 &  0.78133  \\
        \hline
        Two & ensemble + tta & 0.80066 & 0.77539 &  \textbf{0.78802}  \\
        \hline
        Three & +post processing & 0.80066 & 0.77199 &  0.78632  \\
    
        \hline
    \end{tabular}
    \caption{Our submission results on HECKTOR22 leaderboard.}
    \label{tab:result2}
\end{table}

The three submissions we did are:

\begin{itemize}
	\item One - a simple mean ensemble of all models. 
	\item Two - we use Test Time Augmentation (TTA) using axis flips (8 flips total) for each model prediction, which resulted in the best performance.
	\item Three - we attempted to do post-processing on the lymph nodes class based on the submission "Two", by removing small connected components and components with low PET values. Ultimately this heuristic reduced the lymph node accuracy, and was not helpful.   
\end{itemize}

\section{Conclusion}

In conclusion, in this work, we describe our solution to HECKTOR22 challenge (NVAUTO team). Our automated solution is implemented with MONAI\footnote{https://github.com/Project-MONAI/MONAI} and Auto3DSeg\footnote{https://monai.io/apps/auto3dseg}. We achieve the 1st place in the HECKTOR22 challenge segmentation task\footnote{https://hecktor.grand-challenge.org/evaluation/segmentation/leaderboard/}. 

%
\bibliographystyle{splncs04}
\bibliography{mybibliography}

\begin{thebibliography}{1}
\providecommand{\url}[1]{\texttt{#1}}
\providecommand{\urlprefix}{URL }
\providecommand{\doi}[1]{https://doi.org/#1}

\bibitem{monai}
Project-monai/monai, \url{https://doi.org/10.5281/zenodo.5083813}

\bibitem{hecktor22_A}
Andrearczyk, V., Oreiller, V., Boughdad, S., Rest, C.C.L., Elhalawani, H.,
  Jreige, M., Prior, J.O., Vallières, M., Visvikis, D., Hatt, M., Depeursinge,
  A.: Overview of the hecktor challenge at miccai 2022: Automatic head and neck
  tumor segmentation and outcome prediction in pet/ct (2023),
  \url{https://arxiv.org/abs/2201.04138}

\bibitem{he2016identity}
He, K., Zhang, X., Ren, S., Sun, J.: Identity mappings in deep residual
  networks. In: European conference on computer vision. pp. 630--645. Springer
  (2016)

\bibitem{myronenko20183d}
Myronenko, A.: {3D MRI} brain tumor segmentation using autoencoder
  regularization. In: International MICCAI Brainlesion Workshop. pp. 311--320.
  Springer (2018)

\bibitem{hecktor22_B}
Oreiller, V., Andrearczyk, V., Jreige, M., Boughdad, S., Elhalawani, H.,
  Castelli, J., Vallières, M., Zhu, S., Xie, J., Peng, Y., Iantsen, A., Hatt,
  M., Yuan, Y., Ma, J., Yang, X., Rao, C., Pai, S., Ghimire, K., Feng, X.,
  Naser, M.A., Fuller, C.D., Yousefirizi, F., Rahmim, A., Chen, H., Wang, L.,
  Prior, J.O., Depeursinge, A.: Head and neck tumor segmentation in pet/ct: The
  hecktor challenge. Medical Image Analysis  \textbf{77},  102336 (2022)

\end{thebibliography}

\end{document}